\documentclass[prb,preprint,showpacs,amsmath,amssymb]{revtex4-1}

\usepackage[dvipdfmx]{graphicx}
\usepackage{bm}
\usepackage{bibunits}

\newcommand{\be}{\begin{equation}}
\newcommand{\ee}{\end{equation}}
\newcommand{\eq}[1]{Eq.~(\ref{#1})}
\newcommand{\fig}[1]{Fig.~\ref{#1}}
\def\bea{\begin{eqnarray}}
\def\eea{\end{eqnarray}}

\def\bra{\langle}
\def\ket{\rangle}
\def\vq{{\bf q}}

\def\vk{{\bf k}}

\def\vr{{\bf r}}
\def\qp{{\bf q}_{\parallel}}

\begin{document}
\begin{bibunit}[naturemag]

\title{Origin of the high-energy charge excitations observed by resonant inelastic x-ray scattering
in cuprate superconductors} 

\author{Andr\'es Greco$^\dag$,  Hiroyuki Yamase$^*$$^\ddag$, and Mat\'{\i}as Bejas$^\dag$}
\affiliation{
{$^\dag$}Facultad de Ciencias Exactas, Ingenier\'{\i}a y Agrimensura and
Instituto de F\'{\i}sica Rosario (UNR-CONICET),
Av. Pellegrini 250, 2000 Rosario, Argentina\\		
{$^\ddag$}National Institute for Materials Science, Tsukuba 305-0047, Japan
}

\date{\today}

\begin{abstract}
The recent development of x-ray scattering techniques revealed 
the charge-excitation spectrum in high-$T_c$ cuprate superconductors. 
While the presence of a dispersive signal in the high-energy 
charge-excitation spectrum is well accepted in the electron-doped cuprates, its interpretation and 
universality are controversial. Since charge fluctuations are observed ubiquitously in cuprate 
superconductors, the understanding of its origin is a pivotal issue. 
Here, we employ the layered $t$-$J$ model with the long-range Coulomb interaction 
and show that an acoustic-like plasmon mode with a gap  
at in-plane momentum (0,0) captures the major features of the high-energy 
charge excitations.  The high-energy charge excitations, therefore,   
should be a universal feature in cuprate superconductors and are expected also 
in the hole-doped cuprates. 
Acoustic-like plasmons in cuprates have not been recognized yet in experiments. 
We propose several experimental tests to distinguish  
different interpretations of the high-energy charge excitations. 

\vspace{10mm}

\noindent Correspondence to: yamase.hiroyuki@nims.go.jp
\end{abstract}

\maketitle

Recent progress of x-ray scattering techniques 
revealed a short-range charge order in hole-doped cuprates (h-cuprates) 
\cite{ghiringhelli12, chang12, achkar12, blackburn13, blanco-canosa14, comin14,da-silva-neto14, tabis14, gerber15,  chang16,tabis17} 
as well as in electron-doped cuprates  (e-cuprates) \cite{da-silva-neto15,da-silva-neto16,da-silva-neto18}. 
To investigate the energy-resolved charge-excitation spectrum, resonant inelastic x-ray scattering (RIXS) 
is the most powerful tool \cite{hashimoto14,peng16,chaix17}. 
RIXS measurements were performed to explore the high-energy region 
in both e-cuprates \cite{wslee14,ishii14,ishii05} and h-cuprates \cite{ishii17,dellea17}. 
For e-cuprates, Ref.~\onlinecite{wslee14} uncovered that 
the high-energy charge excitations form a steep dispersion around 
$\qp=(0,0)$, where $\qp$ is the in-plane momentum, with an excitation gap about $300$ meV 
at $\qp=(0,0)$. On the other hand, 
Refs.~\onlinecite{ishii14} and \onlinecite{ishii05} measured charge excitations in a wider $\qp$ region
than Ref.~\onlinecite{wslee14}. The spectrum was found to be typically broad and to become 
broader toward the Brillouin zone (BZ) boundary. 
In contrast to Ref.~\onlinecite{wslee14}, 
a gap feature around $\qp=(0,0)$ was not resolved. 
The observed charge excitations 
were thus interpreted differently:  
they may be related to a certain mode near a quantum critical point associated with 
a symmetry-broken state in Ref.~\onlinecite{wslee14}, 
whereas they can be intraband particle-hole excitations with strong incoherent character 
in Refs.~\onlinecite{ishii14} and \onlinecite{ishii05}. 
The situation in h-cuprates is more controversial. Ref.~\onlinecite{ishii17} 
showed that high-energy charge excitations similar to the ones in e-cuprates occur also in h-cuprates, 
while Ref.~\onlinecite{dellea17} emphasized that the high-energy charge 
excitations are a unique feature in e-cuprates and do not occur in h-cuprates.

Recently, a theoretical study of the layered $t$-$J$ model with the long-range Coulomb interaction 
implied that the charge excitation spectrum of cuprates is characterized by 
a dual structure in the energy space \cite{bejas17}. 
The low-energy charge excitations correspond to various types of bond-charge fluctuations 
driven by the exchange term ($J$-term),  whereas the high-energy charge excitations 
are essentially independent of the $J$-term and come from 
usual on-site charge fluctuations. 
In this scenario, the high-energy spectrum is dominated by plasmonic excitations with a finite 
out-of-plane momentum $q_z$. The plasmon mode has a gap at $\qp=(0,0)$ and 
its magnitude is proportional to the interlayer hopping $t_z$ (Ref.~\onlinecite{greco16}). 
This theoretical proposal provides a third idea to understand the high-energy charge excitations. 

Therefore 
three different ideas are proposed for the origin of the high-energy charge excitations: 
i) a certain collective mode near a quantum critical point 
associated with a symmetry-broken  state, which should be specific to e-cuprates \cite{wslee14,dellea17}, 
ii) intraband particle-hole excitations (not plasmons) present in both e- and h-cuprates \cite{ishii14,ishii05,ishii17}, 
and iii) a plasmon mode with a finite $q_z$ (Ref.~\onlinecite{greco16}).  
In this paper, we show that the plasmon scenario yields results consistent with  
the experimental observations. We propose several experimental tests to distinguish different scenarios.

\begin{figure}
\centering
\includegraphics[width=12cm]{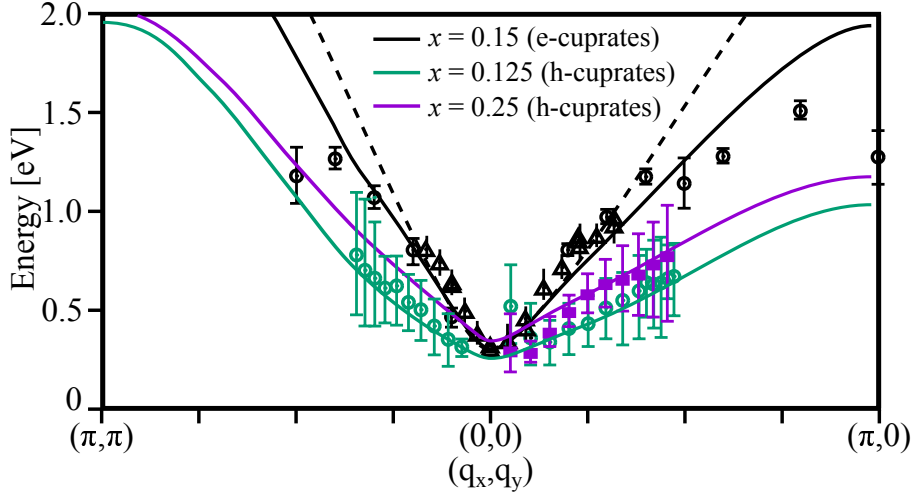}
\caption{
{\bf Plasmon dispersion.}  
Our theoretical results are denoted by the black line for e-cuprates with $x=0.15$, 
and green and purple lines for h-cuprates with $x=0.125$ and $x=0.25$, respectively, 
along the $(\pi,\pi)$-$(0,0)$-$(0,\pi)$ direction.  
For comparison, experimental data are plotted: 
Black up-triangles and black circles for Nd$_{2-x}$Ce$_x$CuO$_4$ with $x=0.15$ 
reported in Ref.~\onlinecite{wslee14} and Refs.~\onlinecite{ishii14} and \onlinecite{ishii05}, 
respectively, and green circles and purple squares for La$_{2-x}$(Br,Sr)$_x$CuO$_4$
with $x=0.125$ and $x=0.25$, respectively \cite{ishii17}.  
The dashed line is the dispersion proposed in Ref.~\onlinecite{wslee14}. 
}
\label{disp}
\end{figure}

We compute the imaginary part of the usual charge susceptibility Im$\chi_c(\vq,\omega)$ 
in the layered $t$-$J$ model with the long-range Coulomb interaction in a large-$N$ scheme 
(see Methods and Supplemental Material).  
We show in \fig{disp} the high-energy collective dispersions, namely 
the peak positions of Im$\chi_c(\vq,\omega)$, 
for e-cuprates with doping $x=0.15$ (black line),
and for h-cuprates with $x=0.125$ (green line)  and $x=0.25$ (purple line). 
These excitations correspond to plasmons realized in the layered system with a gap 
at $\qp=(0,0)$.  
For comparison, we include in \fig{disp} the peak position of the charge excitations obtained in the 
experiments \cite{ishii05,ishii14,ishii17,wslee14}.  
The agreement with the experimental data is very good in $|\qp| \lesssim 0.5\pi$ 
for both e- and h-cuprates. 

Our obtained dispersion for $x=0.25$ has higher energy  
than that for $x=0.125$. This feature captures the experimental 
data shown in \fig{disp}. In fact, the analysis in Ref.~\onlinecite{ishii17} finds that the energy at 
$\qp=(0.46\pi,0)$ increases by a factor of 1.16 when doping is increased from $x=0.125$ to $x=0.25$. 
In the present theory, we obtain a factor of 1.30. 
Similar results were also obtained in the 
density-matrix renormalization-group calculations in the three-band Hubbard model \cite{ishii17}. 

For $x=0.15$ in e-cuprates in the large $\qp$ region, i.e., $(0.5\pi,0)$-$(\pi,0)$, the experimental data 
deviate downward from the dispersion proposed in Ref.~\onlinecite{wslee14} (dashed line in \fig{disp}) 
and tend to be closer to our obtained dispersion. Still, the deviation between the experimental data 
and our results seems substantial, compared with the agreement in the small $\qp$ region. 
However, we think that such a deviation could be related to, as we shall discuss later, the broad spectrum 
observed in the experiments especially in a large $\qp$ region. 

In addition to the agreement with the experimental data in \fig{disp}, the present theory implies 
the following: i) The high-energy charge excitations correspond to a plasmon mode with a gap 
at $\qp=(0,0)$; the gap is proportional to $t_z$ (Ref.~\onlinecite{greco16}). 
ii) Our high-energy charge excitations are present in both e- and h-cuprates. 
iii) The dispersion around $\qp=(0,0)$ has a larger slope in e-cuprates than h-cuprates, 
consistent with the observation in Ref.~\onlinecite{ishii17}. 
iv) The dispersion is rather symmetric between the direction $(0,0)$-$(\pi,0)$ and $(0,0)$-$(\pi,\pi)$.

\begin{figure}
\centering
\includegraphics[width=16cm]{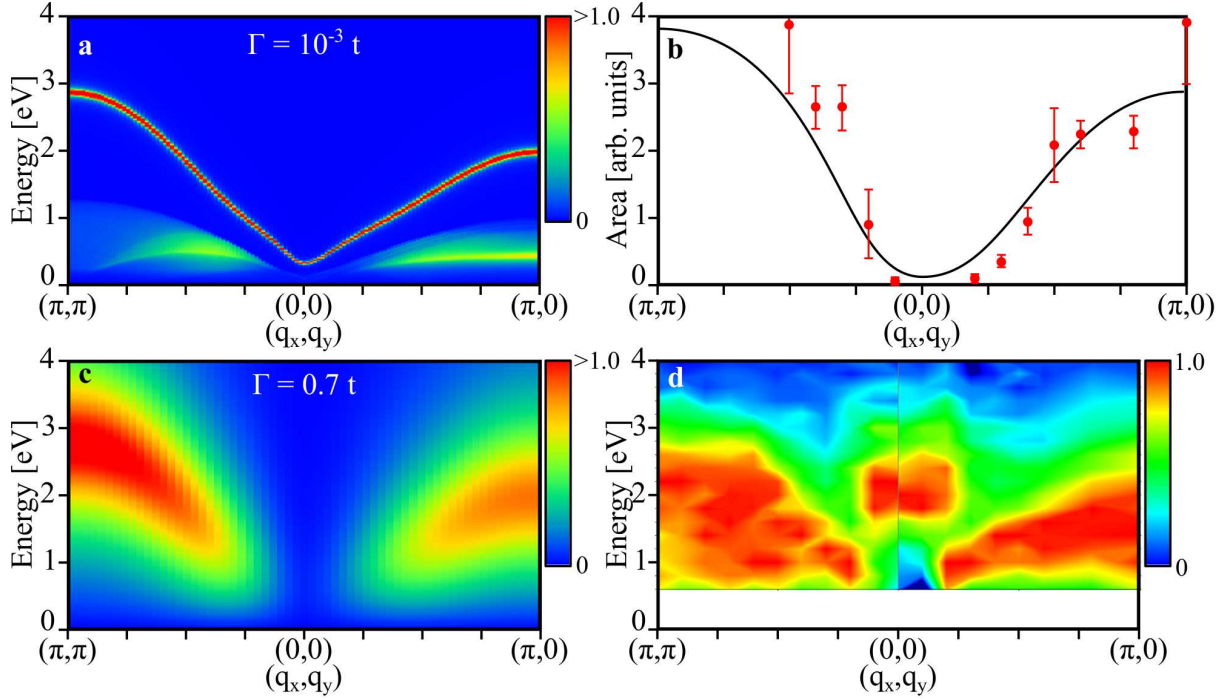}
\caption{
{\bf Charge excitations for e-cuprates with {\boldmath $x=0.15$}.}
{\bf a} and {\bf c},  $\qp$-$\omega$ spectral weight 
for $\Gamma=0.001t$ {\bf a} and $\Gamma=0.7t$ {\bf c}. 
In {\bf a}, the sharp spectrum (red) comes from plamons whereas weak intensity 
around 0.5 eV (green and yellow) is due to individual particle-hole excitations.  
The experimental result of Fig.2(d) in Ref.~\onlinecite{ishii05}
is reproduced in {\bf d} for comparison with {\bf c}. {\bf b}, $\qp$ dependence of the peak area 
obtained from {\bf c}. 
Red circles are the experimental results 
for Nd$_{2-x}$Ce$_x$CuO$_4$ with $x=0.15$ (Ref.~\onlinecite{ishii05}).} 
\label{qw-maps}
\end{figure}

Since the plasmon is a collective mode, it can form a very sharp peak in $\qp$-$\omega$ space 
as shown in \fig{qw-maps}a. On the other hand, 
the experiments of Refs.~\onlinecite{ishii14,ishii05,ishii17} do not show a peak 
signal at $\qp=(0,0)$ and in addition, the spectrum is broad and becomes  
broader with increasing $\qp$ toward the BZ boundary (\fig{qw-maps}d).  
These features are not seen in \fig{qw-maps}a because it was computed in an ideal situation 
by taking $\Gamma=0.001t$, i.e., the damping is assumed to be very small 
(see Methods for the definition of $\Gamma$). 
Considering a realistic situation, we compute the charge-excitation spectrum by employing a large $\Gamma$. 
In principle, $\Gamma$ would depend on momentum and energy, but we take a constant 
$\Gamma=0.7t$ as the simplest case. 
As shown in \fig{qw-maps}c, the spectrum is substantially broadened and becomes broader 
with increasing $\qp$. In addition, the spectrum near $\qp=(0,0)$ becomes poorly 
resolved. These features are very similar to the experimental results (\fig{qw-maps}d). 
In \fig{qw-maps}d, there is strong intensity around $\qp=(0,0)$ and $\omega=2$ eV, which 
comes from the charge transfer excitations between oxygens and coppers, 
namely interband excitations. 
This feature is beyond the scope of the analysis of the present one-band model. 

The inclusion of a large $\Gamma$ is actually invoked theoretically when the spectral line shape 
is compared with experiments \cite{ishii17}. In Ref.~\onlinecite{greco16} a finite $\Gamma$ was also 
used to discuss the temperature dependence of spectral weight \cite{wslee14}.
Physically there should  be two different broadenings, intrinsic and extrinsic ones. 
The extrinsic broadening is due to the instrumental resolution, which is about 250 meV 
($\Gamma \sim 0.35 t$) in Ref.~\onlinecite{ishii14}, 
and 130 meV ($\Gamma \sim 0.2 t$) in Ref.~\onlinecite{wslee14}. 
Because of this difference in  the instrumental resolution, it is possible that the charge excitation 
around $\qp=(0,0)$ is resolved in Ref.~\onlinecite{wslee14}, but not in Ref.~\onlinecite{ishii14}. 
In addition, in the low $\qp$ region studied 
in Ref.~\onlinecite{wslee14} the peak width seems to become broader with increasing $\qp$, 
consistent with Refs.~\onlinecite{ishii14} and \onlinecite{ishii05}. 
The intrinsic broadening comes from incoherent features of high-energy charge excitations 
due to electron correlation effects as demonstrated in various numerical calculations in the $t$-$J$ 
(Refs.~\onlinecite{prelovsek99} and \onlinecite{tohyama95})  
and Hubbard \cite{ishii14,ishii17} models.

We have also calculated the peak area at a given $\qp$ along the $(\pi,\pi)$-$(0,0)$-$(0,\pi)$ direction 
and compare it with the experimental results in \fig{qw-maps}b. 
This agreement with the experiment strengthens the idea that 
the high-energy charge excitations are plasmons.  

\begin{figure}
\centering
\includegraphics[width=16cm]{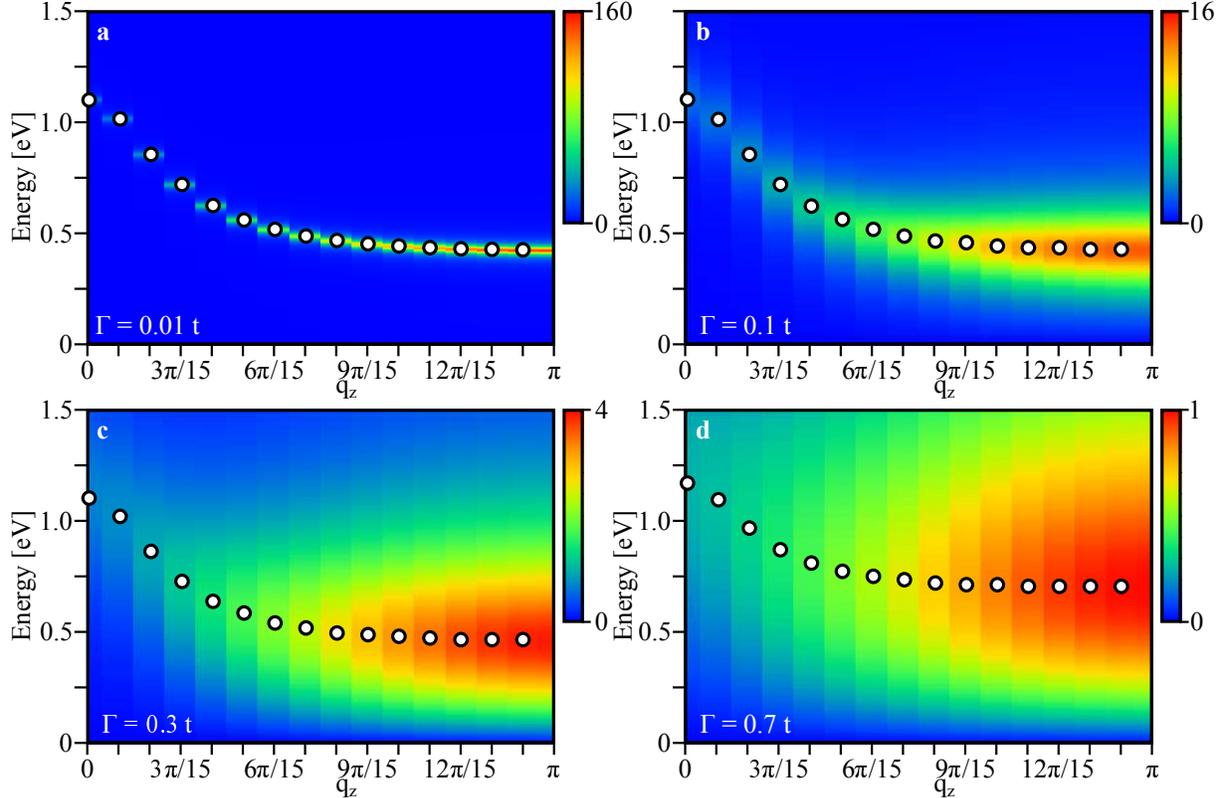}
\caption{
{\bf Intensity maps of plasmons in the plane of {\boldmath $q_z$} and {\boldmath $\omega$.}} 
The in-plane momentum $\qp$ is set to $(0.05\pi, 0.05\pi)$ and the four panels  
correspond to different values of the broadening $\Gamma$. The open circles denotes 
the peak position at each $q_z$. 
}
\label{qzw-maps}
\end{figure}

Our plasmon mode should not be confused with usual optical plasmons, which 
are actually observed in optical measurements \cite{singley01} and 
electron energy-loss spectroscopy \cite{nuecker89,romberg90} in cuprate superconductors. 
The optical plasmon mode is in fact reproduced in our theory by invoking $q_z=0$ (Ref.~\onlinecite{greco16}). 
However, $q_z$ is usually finite in RIXS. Once $q_z$ becomes finite, the optical plasmon energy 
is substantially suppressed to be proportional to the interlayer hopping $t_z$, 
yielding acoustic-like  plasmons as shown in \fig{disp}. 
While this strong $q_z$ dependence was, in part, already discussed in Ref.~\onlinecite{greco16}, as well as 
in early theoretical works where $t_z=0$ was assumed in a layered model \cite{kresin88,bill03,markiewicz08}, 
we present further results. 
Figure~\ref{qzw-maps} shows a map of the spectral weight of plasmons 
in the plane of $q_z$ and $\omega$ for several choices of $\Gamma$ at a small $\qp$.  
The plasmon energy rapidly decreases with increasing $q_z$ and stays almost constant in $q_z > \pi/3$; 
this rapid change is more pronounced when a smaller $\qp$ is chosen. 
The plasmon intensity, on the other hand,  increases with increasing $q_z$, following nearly a 
$q_z^{2}$ dependence at small $q_z$. 
Those qualitative features are independent of the broadening $\Gamma$. 
However, the peak intensity at a small $\qp$ is suppressed substantially with increasing $\Gamma$ 
(see also \fig{qw-maps}c). 
Hence the $q_z$ dependence of plasmons may be 
well observed for a small $\Gamma$. 
Although the importance of the $q_z$ dependence of 
plasmons was not recognized in experimental papers \cite{ishii05,wslee14,ishii14,ishii17,dellea17}, 
we have learned that results similar to \fig{qzw-maps} are recently obtained in 
RIXS (Ref.~\onlinecite{hepting18}).  

The reason why plasmons show a strong $q_z$ dependence is easily understood by recalling 
that plasmons originate from the singularity of the long-range Coulomb interaction in the limit of 
long wavelength. A special feature of the present layered model lies in the anisotropy of the 
momentum dependence of the long-range Coulomb interaction $V(\vq)$ [see \eq{LRC}]. 
When $q_z=0$, $V(\vq)$ is singular at $\qp=(0,0)$, which leads to usual optical plasmons.   
However, due to the anisotropy between $\qp$ and $q_z$, the plasmon energy becomes 
different when $q_z$ is reduced to zero at $\qp=(0,0)$. 
In particular, the plasmon energy would become zero if the interlayer hopping is neglected. 
This is the reason why the plasmon energy becomes 
sensitive to the value of $q_z$, especially in a region of a small $\qp$.  

We have shown that acoustic-like plasmon excitations with a gap at $\qp=(0,0)$ 
due to a finite interlayer hopping describe the main features observed 
by different experimental groups \cite{ishii05,ishii14,ishii17,wslee14} in a consistent way. 
How about other scenarios? 

The scenario proposed in Refs.~\onlinecite{wslee14,dellea17}  invokes 
a collective mode associated with a certain symmetry-broken state to understand 
the high-energy charge excitations around $\qp = (0,0)$. 
The crucial point of this scenario is that their hypothetical order should be specific to 
e-cuprates, because Refs.~\onlinecite{wslee14,dellea17} claim that similar charge 
excitations around $\qp=(0,0)$ are not present in h-cuprates at least 
from their RIXS measurements using the Cu $L_3$-edge. However, both spin and charge 
excitations are detected in the Cu $L_3$-edge and the intensity from the later can be lower than the former. 
In fact, recently Ref.~\onlinecite{ishii17} successfully detects  
the high-energy charge excitations also in h-cuprates by using the oxygen $K$-edge RIXS,  
which can probe directly the charge dynamics of doped holes. 
This recent experimental data is not reconciled with 
the scenario proposed in Refs.~\onlinecite{wslee14,dellea17}. 
Thus, in order to discriminate the scenario of Refs.~\onlinecite{wslee14,dellea17} from the others, 
it is crucial to test the presence of high-energy charge excitations 
in h-cuprates by different experimental groups.

Refs.~\onlinecite{ishii05,ishii14,ishii17} propose intraband charge excitations and 
emphasize their incoherent character. 
This scenario can be interpreted in two different ways. 
On one hand, our acoustic-like plasmons are also from intraband particle-hole excitations 
and their experimental data (Figs.~\ref{disp}, \ref{qw-maps}b, and \ref{qw-maps}d) are well 
captured by introducing a large $\Gamma$ (Figs.~\ref{qw-maps}b and \ref{qw-maps}c).  
Therefore their scenario can be 
understood in terms of plasmons with a large damping,  
although this possibility was not discussed in Refs.~\onlinecite{ishii05,ishii14,ishii17}.  
On the other hand, incoherent intraband excitations occur also as individual charge excitations, 
which form the continuum spectrum below the plasmon energy. 
In this case, two predictions are possible.
First, charge excitations should be gapless at $\qp=(0,0)$. 
Although a region around $\qp=(0,0)$ was not resolved below 2 eV in their measurements 
(see \fig{qw-maps}d), Ref.~\onlinecite{wslee14} reports a gap feature 
(see also \fig{disp}).  Therefore, besides Ref.~\onlinecite{wslee14}, it is decisively important to confirm 
the presence of a charge gap at $\qp=(0,0)$ by different experimental groups. 
The second prediction concerns the $q_z$ dependence of the charge excitations. 
We have computed the $q_z$ dependence of individual charge excitations in our model 
as shown in \fig{qzw-maps-Pi} (see Supplemental Material for details). 
While the peak position depends on choices of $\Gamma$, its $q_z$ dependence 
is almost negligible. This feature is qualitatively different from plasmons shown in \fig{qzw-maps} 
and thus serves to clarify the underlying physics of the high-energy charge excitations.

\begin{figure}
\centering
\includegraphics[width=14cm]{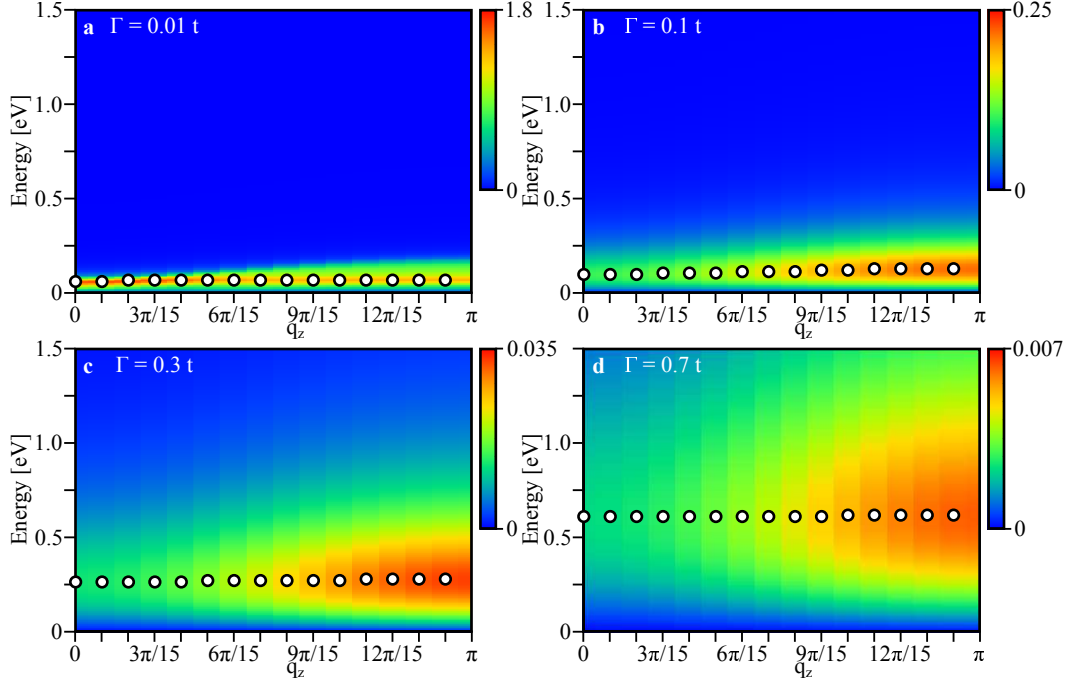}
\caption{
{\bf Intensity maps of individual particle-hole excitations in the plane of {\boldmath $q_z$} and {\boldmath $\omega$}.} 
The parameters are the same as those in \fig{qzw-maps}.  
}
\label{qzw-maps-Pi}
\end{figure}

\begin{figure}
\centering
\includegraphics[width=8cm]{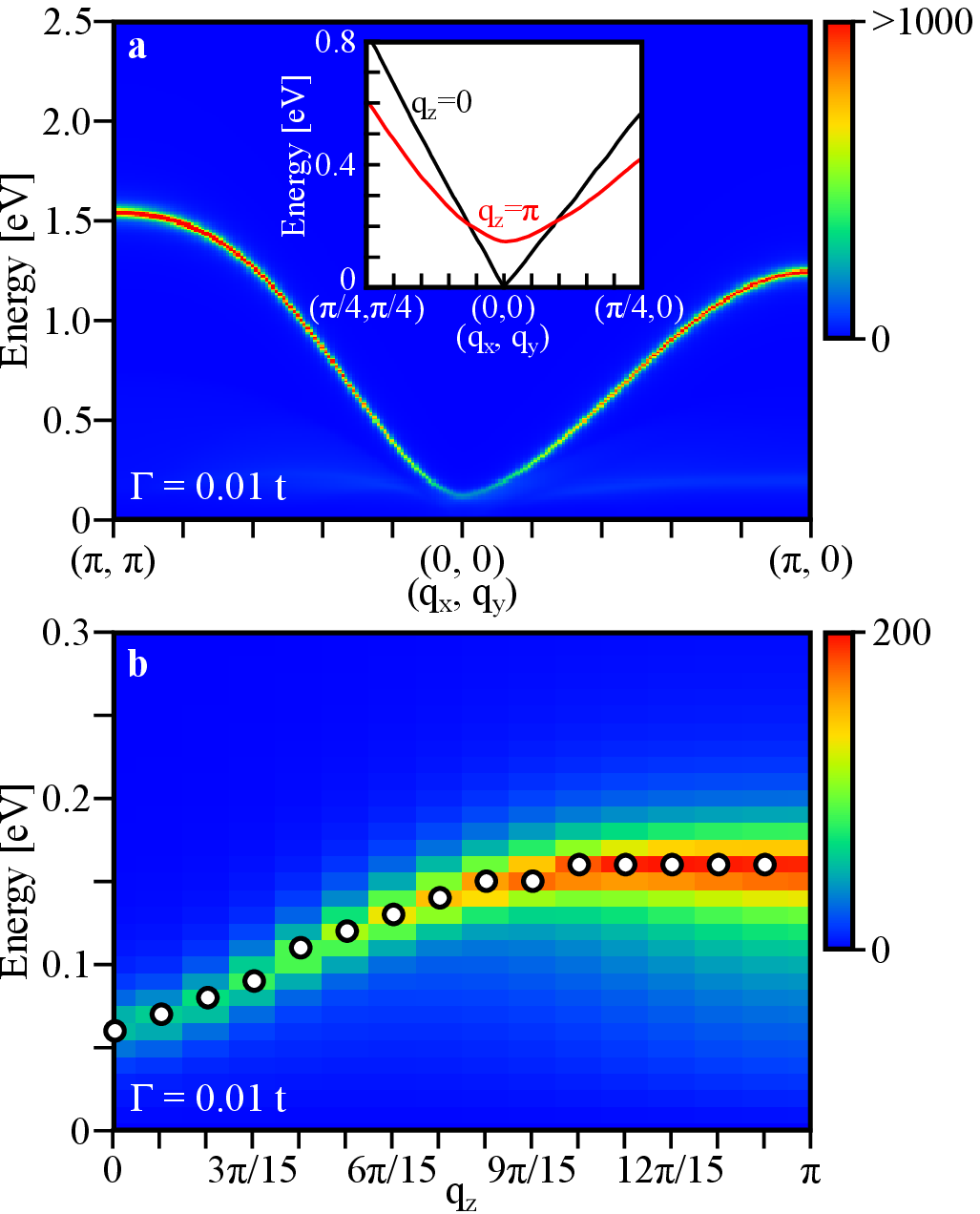}
\caption{
{\bf Intensity maps of charge excitations without the long-range Coulomb interaction.} 
{\bf a}, $\qp$-$\omega$ spectral weight for $q_z=\pi$, a similar plot to \fig{qw-maps}a; 
$\Gamma=0.01t$ is taken for numerical conveneince. 
The sharp spectrum describes a zero-sound mode. {\bf Inset}, zero-sound dispersion for 
$q_z=\pi$ and $0$ around $\qp=(0,0)$. 
{\bf b}, $q_z$ dependence of the intensity at $\qp=(0.02 \pi, 0.02 \pi)$, a similar plot to \fig{qzw-maps}. 
}
\label{zero-sound}
\end{figure}

One may wish to consider a scenario without the long-range Coulomb interaction, which 
may replace plasmons with a zero-sound mode in the $t$-$J$ model.  
To demonstrate this, we have computed charge excitations in our model by using 
the short-range Coulomb interaction instead of the long-range one (see Supplemental Material). 
Our obtained spectrum is shown in \fig{zero-sound}a, which is qualitatively similar to \fig{qw-maps}a. 
While it is clear theoretically that the zero-sound mode is fundamentally different from 
plasmons \cite{negele}, their distinction is less clear from an experimental point of view. 
Hence, we have computed the $q_z$ dependence of the zero-sound mode for a small $\qp$ in \fig{zero-sound}b.  
The zero-sound energy {\it increases} with increasing $q_z$ in a small $q_z$ region, which 
is qualitatively different from the plasmon case shown in \fig{qzw-maps}. 
This is because the zero-sound mode 
becomes gapless at $\qp=(0,0)$ and $q_z=0$ as shown in the inset of \fig{zero-sound}a. 
Therefore, besides Ref.~\onlinecite{hepting18}, 
additional experimental data about the $q_z$ dependence of the high-energy charge excitations 
may confirm the importance of the long-range Coulomb interaction in the charge dynamics in cuprates.

We have demonstrated that acoustic-like plasmon excitations can consistently explain 
experimental data obtained by different groups \cite{ishii05,ishii14,ishii17,wslee14}. 
Conceptually plasmons are well known in solids, but the presence of the acoustic-like 
plasmon mode in cuprates has not been recognized yet in experiments. 
Thus our theoretical recognition of the acoustic-like plasmons in cuprates  
highlights the importance of charge dynamics in cuprates. 
Recalling that cuprates have been studied largely by focusing on spin degrees of freedom, 
it is worth exploring unresolved issues in cuprates 
such as the origin of the pseudogap and the mechanism of high-$T_c$ superconductivity 
in terms of charge degrees of freedom, including the present acoustic-like plasmons.

\vspace{5mm}
{\bf Methods}

We employ the $t$-$J$ model on a square lattice by including the interlayer hopping and 
the long-range Coulomb interaction: 
\begin{equation}
H = -\sum_{i, j,\sigma} t_{i j}\tilde{c}^\dag_{i\sigma}\tilde{c}_{j\sigma} + 
\sum_{\langle i,j \rangle} J_{ij} \left( \vec{S}_i \cdot \vec{S}_j - \frac{1}{4} n_i n_j \right)
+ \frac{1}{2}\sum_{i,j} V_{ij} n_i n_j \, 
\label{tJV}  
\end{equation}
where $\tilde{c}^\dag_{i\sigma}$ and $\tilde{c}_{i\sigma}$ are  
the creation and annihilation operators of electrons 
with spin $\sigma$ 
in the Fock space without double occupancy, and $i$ and $j$ run over a three-dimensional lattice. 
$n_i=\sum_{\sigma} \tilde{c}^\dag_{i\sigma}\tilde{c}_{i\sigma}$ 
is the electron density operator and $\vec{S}_i$ is the spin operator. 
The hopping $t_{i j}$ takes a value $t$ $(t')$ between the first (second) nearest-neighbors 
sites on the square lattice, and $t_z$  between the layers.
$\langle i,j \rangle$ denotes a nearest-neighbor pair of sites. We neglect the magnetic
exchange interaction between the planes, which is much smaller than in-plane $J$ (Ref.~\onlinecite{thio88}).
$V_{ij}$ is the long-range Coulomb interaction on the lattice and it is given in momentum space by \cite{becca96} 
\be
V(\vq)=\frac{V_c}{A(q_x,q_y) - \cos q_z} \,,
\label{LRC}
\ee
where $V_c= e^2 d(2 \epsilon_{\perp} a^2)^{-1}$ and 
\be
A(q_x,q_y)=\alpha (2 - \cos q_x - \cos q_y)+1 \,,
\ee
where $\alpha=(d/a)^2 \epsilon_{\parallel}/\epsilon_{\perp}$,  
and $\epsilon_\parallel$ and $\epsilon_\perp$ are the 
dielectric constants parallel and perpendicular to the planes, respectively; 
$e$ is the electric charge of electrons;  
$a$ is the lattice constant in the planes and the in-plane momentum $\qp=(q_x,q_y)$ is measured 
in units of $a^{-1}$; similarly $d$ is the distance between the planes and the 
out-of-plane momentum $q_z$ is measured in units of $d^{-1}$.  

Since the Hamiltonian (\ref{tJV}) is defined in the Fock space without double occupancy, 
its analysis is not straightforward. Here we use a large-$N$ technique based on  path 
integral representation of the Hubbard $X$ operators \cite{foussats04,bejas12}, which was 
extended to a layered model in Ref.~\onlinecite{greco16}. 
In Supplemental Material, 
we present the essential part of our formalism. 

The model Hamiltonian (\ref{tJV}) contains several parameters, which may 
depend on materials. Fixing a value of $J$ and the long-range Coulomb interaction $V(\vq)$ 
as  $J/t=0.3$ (Ref.~\onlinecite{hybertsen90}),  
and $V_c/t=17$ and $\alpha=4.5$ (Ref.~\onlinecite{greco16}), 
respectively, we  allow a material dependence of the other parameters:  
$t'/t=0.30$ and $t_z/t=0.1$ for Nd$_{2-x}$Ce$_x$CuO$_4$ (Ref.~\onlinecite{greco16}),  and 
$t'/t=-0.20$ and $t_z/t=0.05$ (Ref.~\onlinecite{horio18}) for La$_{2-x}$(Br,Sr)$_x$CuO$_4$. 
It would also be possible to assume a similar value of $t_z$ between them and 
to take a slightly different values of $V_c$ and $\alpha$. 
However, we do not attempt such a tuning. Instead we aim to 
capture the major features of charge-excitation spectrum with a minimal change of the parameters. 
We take the number of layers as $N_z=30$, which should be large enough. 
To address the high-energy charge-excitation spectrum, 
we compute the imaginary part of the usual charge susceptibility 
$\chi_c(\vq,{\rm i}\omega_n)$ 
after  analytical continuation ${\rm i}\omega_n \rightarrow \omega + {\rm i} \Gamma$ 
in Eq.~(S9) in Supplemental Material, 
where $\Gamma$ is positive and, in principle, infinitesimally small. 
In RIXS, the interlayer momentum transfer $q_z$ is usually finite. 
We thus first present results for $q_z=\pi$ as representative ones in Figs.~\ref{disp} and \ref{qw-maps},  
and then study their $q_z$ dependence.   The temperature is set to zero. 

In the comparison with the experimental data in \fig{disp}, 
we have used $t=750$ meV for Nd$_{2-x}$Ce$_x$CuO$_4$
and $t=500$ meV for La$_{2-x}$(Br,Sr)$_x$CuO$_4$. 
The value of $t$ for the former is somewhat larger than the accepted 
value \cite{hybertsen90}. 
A possible reason 
lies in the renormalization of the bare hopping $t$ in the large-$N$ scheme,  
i.e., $t \rightarrow t \frac{x}{2}$  [see Eqs.~(S7) and (S8)  
in Supplemental Material], 
which is simply reduced by $x$ and thus can be simple quantitatively.

\vspace{5mm}
{\bf Author contributions}

A. G. and H. Y. contributed equally to the present project and managed it together. 
H. Y. wrote the major part of the manuscript, and 
A. G. and M. B. performed numerical calculations of the charge susceptibility. 

\vspace{5mm}
{\bf Acknowledgments}

The authors thank T. Devereaux, K. Ishii, and T. Tohyama  for very fruitful discussions, 
and K. Ishii for providing them the experimental data in \fig{qw-maps}b.  
H. Y. acknowledges support by JSPS KAKENHI Grant Number JP15K05189. 
A. G. acknowledges the Japan Society for the Promotion of Science
for a Short-term Invitational Fellowship program (S17027), under which this work was initiated.



\putbib[main_natnew]
\end{bibunit}

\begin{bibunit}[apsrev]
\newpage
\onecolumngrid
\begin{center}
\textbf{\large Supplemental Material: \\
Origin of the high-energy charge excitations observed by resonant inelastic x-ray scattering
in cuprate superconductors} \\
\medskip 

Andr\'es Greco$^\dag$,  Hiroyuki Yamase$^{*}$$^\ddag$, and Mat\'{\i}as Bejas$^\dag$ \\
\smallskip

${}^\dag${\em Facultad de Ciencias Exactas, Ingenier\'{\i}a y Agrimensura and
Instituto de F\'{\i}sica Rosario (UNR-CONICET),
Av. Pellegrini 250, 2000 Rosario, Argentina}

${}^\ddag${\em National Institute for Materials Science, Tsukuba 305-0047, Japan}

\end{center}

\bigskip


In this supplemental material we present i) essential part of our formalism, 
ii) analysis of individual charge excitations, and iii) short-range Coulomb interaction. 
  
\section{Theoretical scheme}
In the path integral formalism \cite{foussats04}, the Hamiltonian (1) can be written in terms 
of an effective model where fermionic fields interact with the six-component 
bosonic field 

\begin{equation}
\delta X^a_{i}=(\delta R_{i}, \delta \lambda_{i}, r^x_{i}, r^y_{i}, A^x_{i}, A^y_{i}). 
\label{bosonX}
\end{equation}

\noindent Here $\delta R_i$ 
describes on-site charge fluctuations and is related to 
$X^{00}_i=N\frac{x}{2} (1+\delta R_i)$ where $X^{00}_i$ is the Hubbard operator \cite{hubbard63} 
associated with the number of holes at a site $i$; 
$x$ is the doped carrier density per site; the factor $N$ comes from 
the sum over the $N$ fermionic channels after the extension of the spin index 
$\sigma$ from $2$ to $N$. 
$\delta \lambda_i$ describes fluctuations of the 
Lagrangian multiplier introduced to impose the constraint of non-double occupancy at any site. 
$r^x_i$ and $r^y_i$ ($A^x_i$ and $A^y_i$) are fluctuations of the real (imaginary) part of 
a bond field along the $x$ and $y$ direction, respectively. 
These bond fields are Hubbard-Stratonovich fields, $\Delta_{i}^{x(y)}$, introduced 
to decouple the exchange interaction in the model (1) and are parametrized as 
$\Delta_i^{x(y)}=\Delta(1+r_i^{x(y)}+i A_i^{x(y)})$, where 
$\Delta$ is the mean-field value of the bond field and is proportional to $J$. 
The bare propagator $D_{ab}^{(0)}(\vq, {\rm i}\omega_n)$ of the bosonic field
$\delta X^{a}$ is given by 
\begin{widetext}
\begin{equation} \label{D0inverse}
[D^{(0)}_{ab}(\vq,\mathrm{i}\omega_n)]^{-1} = N 
\left(
\begin{array}{llllll}
\frac{x^2}{2} \left[ V(\vq)-J(\vq)\right] 
& \frac{x}{2} & 0 & 0 & 0 & 0\\
\frac{x}{2} & 0 & 0 & 0 & 0 & 0\\
0 & 0 & \frac{4\Delta^2}{J} & 0 & 0 & 0\\
0 & 0 & 0 & \frac{4\Delta^2}{J} & 0 & 0\\
0 & 0 & 0 & 0 & \frac{4\Delta^2}{J} & 0\\
0 & 0 & 0 & 0 & 0 & \frac{4\Delta^2}{J}
\end{array}
\right) \; ,
\end{equation}
\end{widetext}
where $J(\vq) = \frac{J}{2} (\cos q_x +  \cos q_y)$ and the matrix indices $a$ and $b$ run from 1 to 6; 
$\vq$ is a three dimensional wavevector and $\omega_n$ is a bosonic Matsubara frequency. 

At leading order, the bare bosonic propagator is renormalized to be 
\be
D^{-1}_{ab}(\vq,\mathrm{i}\omega_n)
= [D^{(0)}_{ab}(\vq,\mathrm{i}\omega_n)]^{-1} - \Pi_{ab}(\vq,\mathrm{i}\omega_n)\,,
\label{dyson}
\ee
where $\Pi_{ab}(\vq,\mathrm{i}\omega_n)$ is the $6\times6$ bosonic self-energy 
\begin{widetext}
\begin{eqnarray}
&& \Pi_{ab}(\vq,\mathrm{i}\omega_n)
            = -\frac{N}{N_s N_z}\sum_{\vk} h_a(\vk,\vq,\varepsilon_\vk-\varepsilon_{\vk-\vq}) 
            \frac{n_F(\varepsilon_{\vk-\vq})-n_F(\varepsilon_\vk)}
                                  {\mathrm{i}\omega_n-\varepsilon_\vk+\varepsilon_{\vk-\vq}} 
            h_b(\vk,\vq,\varepsilon_\vk-\varepsilon_{\vk-\vq}) \nonumber \\
&& \hspace{25mm} - \delta_{a\,1} \delta_{b\,1} \frac{N}{N_s N_z}
                                       \sum_\vk \frac{\varepsilon_\vk-\varepsilon_{\vk-\vq}}{2}n_F(\varepsilon_\vk) \; , 
                                       \label{Pi}
\label{self}
\end{eqnarray}
\end{widetext}
due to the coupling between the bosonic field and fermionic fields; 
$n_F$ is the Fermi-Dirac distribution function. The six components interaction vertex is given by  
\begin{widetext}
\begin{align}
 h_a(\vk,\vq,\nu) =& \left\{
                   \frac{2\varepsilon_{\vk-\vq}+\nu+2\mu}{2}+
                   2\Delta \left[ \cos\left(k_x-\frac{q_x}{2}\right)\cos\left(\frac{q_x}{2}\right) +
                                  \cos\left(k_y-\frac{q_y}{2}\right)\cos\left(\frac{q_y}{2}\right) \right];1;
                 \right. \nonumber \\
               & \left. -2\Delta \cos\left(k_x-\frac{q_x}{2}\right); -2\Delta \cos\left(k_y-\frac{q_y}{2}\right);
                         2\Delta \sin\left(k_x-\frac{q_x}{2}\right);  2\Delta \sin\left(k_y-\frac{q_y}{2}\right)
                 \right\} \, .
\label{vertex-h}
\end{align}
\end{widetext} 
The electronic dispersion $\epsilon_{\vk}$ is defined as 
\be
\varepsilon_{\vk} = \varepsilon_{\vk}^{\parallel}  + \varepsilon_{\vk}^{\perp} \,,
\label{Ek}
\ee
where the in-plane dispersion $\varepsilon_{\vk}^{\parallel}$ and the out-of-plane dispersion 
$\varepsilon_{\vk}^{\perp}$ are given, respectively, by
\be
\varepsilon_{\vk}^{\parallel} = -2 \left( t \frac{x}{2}+\Delta \right) (\cos k_{x}+\cos k_{y})-
4t' \frac{x}{2} \cos k_{x} \cos k_{y} - \mu \,,\\
\label{Epara}
\ee
\be
\varepsilon_{\vk}^{\perp} = 2 t_{z} \frac{x}{2} (\cos k_x-\cos k_y)^2 \cos k_{z}  \,,
\label{Eperp}
\ee
and $\mu$ is the chemical potential. 
Note that the bare hopping integrals $t$, $t'$, and $t_z$ are renormalized by a factor 
$x/2$. No incoherent self-energy effects enter the fermionic dispersion 
at leading order.
Note that $k_z$ and $q_z$ dependences enter only through $\epsilon_{\vk-\vq}$  in the first column 
in \eq{vertex-h}, whereas the other columns contain only the in-plane momentum $\vq_{\parallel}$.
In \eq{self}, $N_s$ and $N_z$ are the total 
number of lattice sites on the square lattice and the number of layers along the $z$ direction, respectively. 

All possible charge excitations in the layered $t$-$J$ model are contained 
in $D_{ab}(\vq,\mathrm{i}\omega_n)$ [\eq{dyson}] and can be treated on equal footing 
in the present theoretical scheme \cite{bejas17}. 
Usual charge fluctuations, namely 
$\chi_c (\vr_i -\vr_j, \tau)=\bra T_\tau n_i(\tau) n_j(0)\ket$, 
are associated with the element $(1,1)$ of the full $6\times6$ $D_{ab}$ in \eq{dyson} and 
is computed in $\qp$-$\omega$ space as \cite{bejas12,bejas17}  
\begin{eqnarray}
\chi_{c}(\vq,{\rm i}\omega_n)= - N \left ( \frac{x}{2} \right )^{2} D_{11}(\vq,{\rm i}\omega_n)  \,.
\label{chic}
\end{eqnarray}
On the other hand, the elements from $3$ to $6$ of the matrix $D_{ab}$ 
describe low-energy charge excitations associated with 
the charge order phenomenon \cite{bejas17}.
Since we are interested in the high-energy charge excitations, we focus here on \eq{chic}.

\section{Individual charge excitations}
The charge susceptibility $\chi_c (\vq, {\rm i}\omega_n)$ [\eq{chic}] 
contains both collective and individual excitations. 
It is not trivial to separate those excitations in a strong coupling model 
such as the $t$-$J$ model.  
To extract the individual excitations,  we consider a diagramatic structure in the 
present large-$N$ scheme and 
focus on contribution from a single bubble diagram. 
This single bubble is given by $\Pi_{22}(\vq,{\rm i}\omega_n)$. 
Intuitively this is clear because the vertex $h_a$ of the single bubble should be constant for 
usual charge correlations, which is fulfilled only for the $a=2$ component in \eq{vertex-h}. 
Mathematically we can show that the individual excitations are described by 
$\Pi_{22}$ at leading order of the large-$N$ scheme when we rewrite  
\eq{chic} in terms of Hubbard $X$-operators $X^{p0}_{i}$ and $X^{0p}_{i}$ by using 
the two constraints such as $X_{i}^{00} + \sum_{p} X_{i}^{pp} = \frac{N}{2}$ and  
$X_{i}^{pp'} = \frac{X_{i}^{p0} X_{i}^{0p'}}{X_{i}^{00}}$ (see Ref. \onlinecite{bejas12}). 
Therefore after the analytical continuation, we have computed Im$\Pi_{22}(\vq,\omega)$ 
in Fig.~4. We have checked that the spectrum of Im$\Pi_{22}(\vq,\omega)$ 
for a small $\Gamma$ is indeed similar to 
the continuum spectrum of  Im$\chi_c (\vq, \omega)$ obtained in Fig.~2a,  
where the plasmon mode   
is well separated from the continuum spectrum.

\section{Short-range Coulomb interaction} 
As a typical short-range Coulomb interaction, we may take 
\be
V(\vq)= V_1 (\cos q_x + \cos q_y ) +V_2 \cos q_z  \,. 
\ee
The charge excitation spectrum for $V_2=0$ and $t_z=0$ was already shown in 
Ref. \onlinecite{greco17}, where a zero-sound mode is realized as collective excitations, 
which are gapless at $\qp=(0,0)$; see Ref. \onlinecite{greco17} for more details 
of the zero-sound mode in the $t$-$J$ model.  
In the present layered model with a finite $t_z$, however, the zero-sound mode acquires  a gap 
at $\qp=(0,0)$ for a finite $q_z$ as shown in Fig.~5a. 
In Fig.~5 we have chosen $V_1=V_2= t$ after checking 
that qualitatively the same results are obtained for 
other choices of $V_1$ and $V_2$ as long as the system is stable in the presence of $V(\vq)$. 



\begin{thebibliography}{10}
\expandafter\ifx\csname url\endcsname\relax
  \def\url#1{\texttt{#1}}\fi
\expandafter\ifx\csname urlprefix\endcsname\relax\def\urlprefix{URL }\fi
\providecommand{\bibinfo}[2]{#2}
\providecommand{\eprint}[2][]{\url{#2}}

\bibitem{ghiringhelli12}
\bibinfo{author}{Ghiringhelli, G.} \emph{et~al.}
\newblock \bibinfo{title}{{Long-Range Incommensurate Charge Fluctuations in
  (Y,Nd)Ba$_2$Cu$_3$O$_{6+x}$}}.
\newblock \emph{\bibinfo{journal}{Science}} \textbf{\bibinfo{volume}{337}},
  \bibinfo{pages}{821--825} (\bibinfo{year}{2012}).
\newblock \urlprefix\url{http://science.sciencemag.org/content/337/6096/821}.

\bibitem{chang12}
\bibinfo{author}{Chang, J.} \emph{et~al.}
\newblock \bibinfo{title}{{Direct observation of competition between
  superconductivity and charge density wave order in YBa$_2$Cu$_3$O$_{6.67}$}}.
\newblock \emph{\bibinfo{journal}{Nature Physics}}
  \textbf{\bibinfo{volume}{8}}, \bibinfo{pages}{871} (\bibinfo{year}{2012}).
\newblock \urlprefix\url{http://dx.doi.org/10.1038/nphys2456}.

\bibitem{achkar12}
\bibinfo{author}{Achkar, A.~J.} \emph{et~al.}
\newblock \bibinfo{title}{{Distinct Charge Orders in the Planes and Chains of
  Ortho-III-Ordered
  ${\rm{YBa}}_{2}{\rm{Cu}}_{3}{\rm{O}}_{6+\ensuremath{\delta}}$ Superconductors
  Identified by Resonant Elastic X-ray Scattering}}.
\newblock \emph{\bibinfo{journal}{Phys. Rev. Lett.}}
  \textbf{\bibinfo{volume}{109}}, \bibinfo{pages}{167001}
  (\bibinfo{year}{2012}).
\newblock
  \urlprefix\url{https://link.aps.org/doi/10.1103/PhfysRevLett.109.167001}.

\bibitem{blackburn13}
\bibinfo{author}{Blackburn, E.} \emph{et~al.}
\newblock \bibinfo{title}{{X-Ray Diffraction Observations of a
  Charge-Density-Wave Order in Superconducting Ortho-II
  ${\rm{YBa}}_{2}{\rm{Cu}}_{3}{\rm{O}}_{6.54}$ Single Crystals in Zero Magnetic
  Field}}.
\newblock \emph{\bibinfo{journal}{Phys. Rev. Lett.}}
  \textbf{\bibinfo{volume}{110}}, \bibinfo{pages}{137004}
  (\bibinfo{year}{2013}).
\newblock
  \urlprefix\url{https://link.aps.org/doi/10.1103/PhysRevLett.110.137004}.

\bibitem{blanco-canosa14}
\bibinfo{author}{Blanco-Canosa, S.} \emph{et~al.}
\newblock \bibinfo{title}{{Resonant x-ray scattering study of charge-density
  wave correlations in ${\rm{YBa}}_{2}{\rm{Cu}}_{3}{\rm{O}}_{6+x}$}}.
\newblock \emph{\bibinfo{journal}{Phys. Rev. B}} \textbf{\bibinfo{volume}{90}},
  \bibinfo{pages}{054513} (\bibinfo{year}{2014}).
\newblock \urlprefix\url{https://link.aps.org/doi/10.1103/PhysRevB.90.054513}.

\bibitem{comin14}
\bibinfo{author}{Comin, R.} \emph{et~al.}
\newblock \bibinfo{title}{{Charge Order Driven by Fermi-Arc Instability in
  Bi$_2$Sr$_{2-x}$La$_x$CuO$_{6+\delta}$}}.
\newblock \emph{\bibinfo{journal}{Science}} \textbf{\bibinfo{volume}{343}},
  \bibinfo{pages}{390--392} (\bibinfo{year}{2014}).
\newblock \urlprefix\url{http://science.sciencemag.org/content/343/6169/390}.

\bibitem{da-silva-neto14}
\bibinfo{author}{da~Silva~Neto, E.~H.} \emph{et~al.}
\newblock \bibinfo{title}{Ubiquitous interplay between charge ordering and
  high-temperature superconductivity in cuprates}.
\newblock \emph{\bibinfo{journal}{Science}} \textbf{\bibinfo{volume}{343}},
  \bibinfo{pages}{393--396} (\bibinfo{year}{2014}).
\newblock \urlprefix\url{http://science.sciencemag.org/content/343/6169/393}.

\bibitem{tabis14}
\bibinfo{author}{Tabis, W.} \emph{et~al.}
\newblock \bibinfo{title}{{Charge order and its connection with Fermi-liquid
  charge transport in a pristine high-$T_c$ cuprate}}.
\newblock \emph{\bibinfo{journal}{Nature Communications}}
  \textbf{\bibinfo{volume}{5}}, \bibinfo{pages}{5875} (\bibinfo{year}{2014}).
\newblock \urlprefix\url{http://dx.doi.org/10.1038/ncomms6875}.

\bibitem{gerber15}
\bibinfo{author}{Gerber, S.} \emph{et~al.}
\newblock \bibinfo{title}{{Three-dimensional charge density wave order in
  YBa$_2$Cu$_3$O$_{6.67}$ at high magnetic fields}}.
\newblock \emph{\bibinfo{journal}{Science}} \textbf{\bibinfo{volume}{350}},
  \bibinfo{pages}{949--952} (\bibinfo{year}{2015}).
\newblock \urlprefix\url{http://science.sciencemag.org/content/350/6263/949}.

\bibitem{chang16}
\bibinfo{author}{Chang, J.} \emph{et~al.}
\newblock \bibinfo{title}{{Magnetic field controlled charge density wave
  coupling in underdoped YBa$_2$Cu$_3$O$_{6+x}$}}.
\newblock \emph{\bibinfo{journal}{Nature Communications}}
  \textbf{\bibinfo{volume}{7}}, \bibinfo{pages}{11494} (\bibinfo{year}{2016}).
\newblock \urlprefix\url{http://dx.doi.org/10.1038/ncomms11494}.

\bibitem{tabis17}
\bibinfo{author}{Tabis, W.} \emph{et~al.}
\newblock \bibinfo{title}{{Synchrotron x-ray scattering study of
  charge-density-wave order in
  ${\mathrm{HgBa}}_{2}{\mathrm{CuO}}_{4+\ensuremath{\delta}}$}}.
\newblock \emph{\bibinfo{journal}{Phys. Rev. B}} \textbf{\bibinfo{volume}{96}},
  \bibinfo{pages}{134510} (\bibinfo{year}{2017}).
\newblock \urlprefix\url{https://link.aps.org/doi/10.1103/PhysRevB.96.134510}.

\bibitem{da-silva-neto15}
\bibinfo{author}{da~Silva~Neto, E.~H.} \emph{et~al.}
\newblock \bibinfo{title}{{Charge ordering in the electron-doped superconductor
  Nd$_{2-x}$Ce$_x$CuO$_4$}}.
\newblock \emph{\bibinfo{journal}{Science}} \textbf{\bibinfo{volume}{347}},
  \bibinfo{pages}{282--285} (\bibinfo{year}{2015}).
\newblock \urlprefix\url{http://science.sciencemag.org/content/347/6219/282}.

\bibitem{da-silva-neto16}
\bibinfo{author}{da~Silva~Neto, E.~H.} \emph{et~al.}
\newblock \bibinfo{title}{Doping-dependent charge order correlations in
  electron-doped cuprates}.
\newblock \emph{\bibinfo{journal}{Science Advances}}
  \textbf{\bibinfo{volume}{2}}, \bibinfo{pages}{e1600782}
  (\bibinfo{year}{2016}).
\newblock \urlprefix\url{http://advances.sciencemag.org/content/2/8/e1600782}.

\bibitem{da-silva-neto18}
\bibinfo{author}{{da Silva Neto}, E.~H.} \emph{et~al.}
\newblock \bibinfo{title}{{{Coupling between dynamic magnetic and charge-order
  correlations in the cuprate superconductor Nd$_{2-x}$Ce$_{x}$CuO$_4$}}}.
\newblock \emph{\bibinfo{journal}{ArXiv e-prints}}  (\bibinfo{year}{2018}).
\newblock \eprint{1804.09185}.

\bibitem{hashimoto14}
\bibinfo{author}{Hashimoto, M.} \emph{et~al.}
\newblock \bibinfo{title}{{Direct observation of bulk charge modulations in
  optimally doped
  ${\mathrm{Bi}}_{1.5}{\mathrm{Pb}}_{0.6}{\mathrm{Sr}}_{1.54}{\mathrm{CaCu}}_{2}{\mathrm{O}}_{8+\ensuremath{\delta}}$}}.
\newblock \emph{\bibinfo{journal}{Phys. Rev. B}} \textbf{\bibinfo{volume}{89}},
  \bibinfo{pages}{220511} (\bibinfo{year}{2014}).
\newblock \urlprefix\url{https://link.aps.org/doi/10.1103/PhysRevB.89.220511}.

\bibitem{peng16}
\bibinfo{author}{Peng, Y.~Y.} \emph{et~al.}
\newblock \bibinfo{title}{{Direct observation of charge order in underdoped and
  optimally doped
  ${\mathrm{Bi}}_{2}{(\mathrm{Sr},\mathrm{La})}_{2}{\mathrm{CuO}}_{6+\ensuremath{\delta}}$
  by resonant inelastic x-ray scattering}}.
\newblock \emph{\bibinfo{journal}{Phys. Rev. B}} \textbf{\bibinfo{volume}{94}},
  \bibinfo{pages}{184511} (\bibinfo{year}{2016}).
\newblock \urlprefix\url{https://link.aps.org/doi/10.1103/PhysRevB.94.184511}.

\bibitem{chaix17}
\bibinfo{author}{Chaix, L.} \emph{et~al.}
\newblock \bibinfo{title}{{Dispersive charge density wave excitations in
  Bi$_2$Sr$_2$CaCu$_2$O$_{8+\delta}$}}.
\newblock \emph{\bibinfo{journal}{Nature Physics}}
  \textbf{\bibinfo{volume}{13}}, \bibinfo{pages}{952} (\bibinfo{year}{2017}).
\newblock \urlprefix\url{http://dx.doi.org/10.1038/nphys4157}.

\bibitem{wslee14}
\bibinfo{author}{Lee, W.~S.} \emph{et~al.}
\newblock \bibinfo{title}{Asymmetry of collective excitations in electron- and
  hole-doped cuprate superconductors}.
\newblock \emph{\bibinfo{journal}{Nature Physics}}
  \textbf{\bibinfo{volume}{10}}, \bibinfo{pages}{883} (\bibinfo{year}{2014}).
\newblock \urlprefix\url{http://dx.doi.org/10.1038/nphys3117}.

\bibitem{ishii14}
\bibinfo{author}{Ishii, K.} \emph{et~al.}
\newblock \bibinfo{title}{High-energy spin and charge excitations in
  electron-doped copper oxide superconductors}.
\newblock \emph{\bibinfo{journal}{Nature Communications}}
  \textbf{\bibinfo{volume}{5}}, \bibinfo{pages}{3714} (\bibinfo{year}{2014}).
\newblock \urlprefix\url{http://dx.doi.org/10.1038/ncomms4714}.

\bibitem{ishii05}
\bibinfo{author}{Ishii, K.} \emph{et~al.}
\newblock \bibinfo{title}{{Momentum Dependence of Charge Excitations in the
  Electron-Doped Superconductor
  ${\mathrm{Nd}}_{1.85}{\mathrm{Ce}}_{0.15}{\mathrm{CuO}}_{4}$: A Resonant
  Inelastic X-Ray Scattering Study}}.
\newblock \emph{\bibinfo{journal}{Phys. Rev. Lett.}}
  \textbf{\bibinfo{volume}{94}}, \bibinfo{pages}{207003}
  (\bibinfo{year}{2005}).
\newblock
  \urlprefix\url{https://link.aps.org/doi/10.1103/PhysRevLett.94.207003}.

\bibitem{ishii17}
\bibinfo{author}{Ishii, K.} \emph{et~al.}
\newblock \bibinfo{title}{{Observation of momentum-dependent charge excitations
  in hole-doped cuprates using resonant inelastic x-ray scattering at the
  oxygen $K$ edge}}.
\newblock \emph{\bibinfo{journal}{Phys. Rev. B}} \textbf{\bibinfo{volume}{96}},
  \bibinfo{pages}{115148} (\bibinfo{year}{2017}).
\newblock \urlprefix\url{https://link.aps.org/doi/10.1103/PhysRevB.96.115148}.

\bibitem{dellea17}
\bibinfo{author}{Dellea, G.} \emph{et~al.}
\newblock \bibinfo{title}{Spin and charge excitations in artificial hole- and
  electron-doped infinite layer cuprate superconductors}.
\newblock \emph{\bibinfo{journal}{Phys. Rev. B}} \textbf{\bibinfo{volume}{96}},
  \bibinfo{pages}{115117} (\bibinfo{year}{2017}).
\newblock \urlprefix\url{https://link.aps.org/doi/10.1103/PhysRevB.96.115117}.

\bibitem{bejas17}
\bibinfo{author}{Bejas, M.}, \bibinfo{author}{Yamase, H.} \&
  \bibinfo{author}{Greco, A.}
\newblock \bibinfo{title}{Dual structure in the charge excitation spectrum of
  electron-doped cuprates}.
\newblock \emph{\bibinfo{journal}{Phys. Rev. B}} \textbf{\bibinfo{volume}{96}},
  \bibinfo{pages}{214513} (\bibinfo{year}{2017}).
\newblock \urlprefix\url{https://link.aps.org/doi/10.1103/PhysRevB.96.214513}.

\bibitem{greco16}
\bibinfo{author}{Greco, A.}, \bibinfo{author}{Yamase, H.} \&
  \bibinfo{author}{Bejas, M.}
\newblock \bibinfo{title}{{Plasmon excitations in layered high-${T}_{c}$
  cuprates}}.
\newblock \emph{\bibinfo{journal}{Phys. Rev. B}} \textbf{\bibinfo{volume}{94}},
  \bibinfo{pages}{075139} (\bibinfo{year}{2016}).
\newblock \urlprefix\url{https://link.aps.org/doi/10.1103/PhysRevB.94.075139}.

\bibitem{prelovsek99}
\bibinfo{author}{Prelov\ifmmode~\check{s}\else \v{s}\fi{}ek, P.} \&
  \bibinfo{author}{Horsch, P.}
\newblock \bibinfo{title}{Electron-energy loss spectra and plasmon resonance in
  cuprates}.
\newblock \emph{\bibinfo{journal}{Phys. Rev. B}} \textbf{\bibinfo{volume}{60}},
  \bibinfo{pages}{R3735--R3738} (\bibinfo{year}{1999}).
\newblock \urlprefix\url{https://link.aps.org/doi/10.1103/PhysRevB.60.R3735}.

\bibitem{tohyama95}
\bibinfo{author}{Tohyama, T.}, \bibinfo{author}{Horsch, P.} \&
  \bibinfo{author}{Maekawa, S.}
\newblock \bibinfo{title}{{Spin and Charge Dynamics of the $t\ensuremath{-}J$
  Model}}.
\newblock \emph{\bibinfo{journal}{Phys. Rev. Lett.}}
  \textbf{\bibinfo{volume}{74}}, \bibinfo{pages}{980--983}
  (\bibinfo{year}{1995}).
\newblock \urlprefix\url{https://link.aps.org/doi/10.1103/PhysRevLett.74.980}.

\bibitem{singley01}
\bibinfo{author}{Singley, E.~J.}, \bibinfo{author}{Basov, D.~N.},
  \bibinfo{author}{Kurahashi, K.}, \bibinfo{author}{Uefuji, T.} \&
  \bibinfo{author}{Yamada, K.}
\newblock \bibinfo{title}{{Electron dynamics in
  Nd$_{1.85}$Ce$_{0.15}$CuO$_{4+\delta}$: Evidence for the pseudogap state and
  unconventional c-axis response}}.
\newblock \emph{\bibinfo{journal}{Phys. Rev. B}} \textbf{\bibinfo{volume}{64}},
  \bibinfo{pages}{224503} (\bibinfo{year}{2001}).
\newblock \urlprefix\url{https://link.aps.org/doi/10.1103/PhysRevB.64.224503}.

\bibitem{nuecker89}
\bibinfo{author}{N\"ucker, N.} \emph{et~al.}
\newblock \bibinfo{title}{{Plasmons and interband transitions in
  ${\mathrm{Bi}}_{2}$${\mathrm{Sr}}_{2}$Ca${\mathrm{Cu}}_{2}$${\mathrm{O}}_{8}$}}.
\newblock \emph{\bibinfo{journal}{Phys. Rev. B}} \textbf{\bibinfo{volume}{39}},
  \bibinfo{pages}{12379--12382} (\bibinfo{year}{1989}).
\newblock \urlprefix\url{https://link.aps.org/doi/10.1103/PhysRevB.39.12379}.

\bibitem{romberg90}
\bibinfo{author}{Romberg, H.} \emph{et~al.}
\newblock \bibinfo{title}{{Dielectric function of YBa$_2$Cu$_3$O$_{7-\delta}$
  between 50 meV and 50 eV}}.
\newblock \emph{\bibinfo{journal}{Zeitschrift f{\"u}r Physik B Condensed
  Matter}} \textbf{\bibinfo{volume}{78}}, \bibinfo{pages}{367--380}
  (\bibinfo{year}{1990}).
\newblock \urlprefix\url{https://doi.org/10.1007/BF01313317}.

\bibitem{kresin88}
\bibinfo{author}{Kresin, V.~Z.} \& \bibinfo{author}{Morawitz, H.}
\newblock \bibinfo{title}{Layer plasmons and high-${T}_{c}$ superconductivity}.
\newblock \emph{\bibinfo{journal}{Phys. Rev. B}} \textbf{\bibinfo{volume}{37}},
  \bibinfo{pages}{7854--7857} (\bibinfo{year}{1988}).
\newblock \urlprefix\url{https://link.aps.org/doi/10.1103/PhysRevB.37.7854}.

\bibitem{bill03}
\bibinfo{author}{Bill, A.}, \bibinfo{author}{Morawitz, H.} \&
  \bibinfo{author}{Kresin, V.~Z.}
\newblock \bibinfo{title}{Electronic collective modes and superconductivity in
  layered conductors}.
\newblock \emph{\bibinfo{journal}{Phys. Rev. B}} \textbf{\bibinfo{volume}{68}},
  \bibinfo{pages}{144519} (\bibinfo{year}{2003}).
\newblock \urlprefix\url{https://link.aps.org/doi/10.1103/PhysRevB.68.144519}.

\bibitem{markiewicz08}
\bibinfo{author}{Markiewicz, R.~S.}, \bibinfo{author}{Hasan, M.~Z.} \&
  \bibinfo{author}{Bansil, A.}
\newblock \bibinfo{title}{Acoustic plasmons and doping evolution of mott
  physics in resonant inelastic x-ray scattering from cuprate superconductors}.
\newblock \emph{\bibinfo{journal}{Phys. Rev. B}} \textbf{\bibinfo{volume}{77}},
  \bibinfo{pages}{094518} (\bibinfo{year}{2008}).
\newblock \urlprefix\url{https://link.aps.org/doi/10.1103/PhysRevB.77.094518}.

\bibitem{hepting18}
\bibinfo{author}{Hepting, M.} \emph{et~al.}
\newblock \bibinfo{title}{Three-dimensional collective charge excitations in
  electron-doped copper oxide superconductors}.
\newblock \emph{\bibinfo{journal}{Nature}}  (\bibinfo{year}{2018}).
\newblock \urlprefix\url{https://doi.org/10.1038/s41586-018-0648-3}.

\bibitem{negele}
\bibinfo{author}{Negele, J.~W.} \& \bibinfo{author}{Orland, H.}
\newblock \emph{\bibinfo{title}{Quantum Many-Particle Systems}}
  (\bibinfo{publisher}{Perseus Books Publishing}, \bibinfo{year}{1998}).

\bibitem{thio88}
\bibinfo{author}{Thio, T.} \emph{et~al.}
\newblock \bibinfo{title}{{Antisymmetric exchange and its influence on the
  magnetic structure and conductivity of
  ${\mathrm{La}}_{2}$Cu${\mathrm{O}}_{4}$}}.
\newblock \emph{\bibinfo{journal}{Phys. Rev. B}} \textbf{\bibinfo{volume}{38}},
  \bibinfo{pages}{905--908} (\bibinfo{year}{1988}).
\newblock \urlprefix\url{https://link.aps.org/doi/10.1103/PhysRevB.38.905}.

\bibitem{becca96}
\bibinfo{author}{Becca, F.}, \bibinfo{author}{Tarquini, M.},
  \bibinfo{author}{Grilli, M.} \& \bibinfo{author}{Di~Castro, C.}
\newblock \bibinfo{title}{{Charge-density waves and superconductivity as an
  alternative to phase separation in the infinite-$U$ Hubbard-Holstein model}}.
\newblock \emph{\bibinfo{journal}{Phys. Rev. B}} \textbf{\bibinfo{volume}{54}},
  \bibinfo{pages}{12443--12457} (\bibinfo{year}{1996}).
\newblock \urlprefix\url{https://link.aps.org/doi/10.1103/PhysRevB.54.12443}.

\bibitem{foussats04}
\bibinfo{author}{Foussats, A.} \& \bibinfo{author}{Greco, A.}
\newblock \bibinfo{title}{{Large-$N$ expansion based on the Hubbard operator
  path integral representation and its application to the
  $t\text{\ensuremath{-}}J$ model. II. The case for finite $J$}}.
\newblock \emph{\bibinfo{journal}{Phys. Rev. B}} \textbf{\bibinfo{volume}{70}},
  \bibinfo{pages}{205123} (\bibinfo{year}{2004}).
\newblock \urlprefix\url{https://link.aps.org/doi/10.1103/PhysRevB.70.205123}.

\bibitem{bejas12}
\bibinfo{author}{Bejas, M.}, \bibinfo{author}{Greco, A.} \&
  \bibinfo{author}{Yamase, H.}
\newblock \bibinfo{title}{Possible charge instabilities in two-dimensional
  doped mott insulators}.
\newblock \emph{\bibinfo{journal}{Phys. Rev. B}} \textbf{\bibinfo{volume}{86}},
  \bibinfo{pages}{224509} (\bibinfo{year}{2012}).
\newblock \urlprefix\url{https://link.aps.org/doi/10.1103/PhysRevB.86.224509}.

\bibitem{hybertsen90}
\bibinfo{author}{Hybertsen, M.~S.}, \bibinfo{author}{Stechel, E.~B.},
  \bibinfo{author}{Schluter, M.} \& \bibinfo{author}{Jennison, D.~R.}
\newblock \bibinfo{title}{{Renormalization from density-functional theory to
  strong-coupling models for electronic states in Cu-O materials}}.
\newblock \emph{\bibinfo{journal}{Phys. Rev. B}} \textbf{\bibinfo{volume}{41}},
  \bibinfo{pages}{11068--11072} (\bibinfo{year}{1990}).
\newblock \urlprefix\url{https://link.aps.org/doi/10.1103/PhysRevB.41.11068}.

\bibitem{horio18}
\bibinfo{author}{Horio, M.} \emph{et~al.}
\newblock \bibinfo{title}{Three-dimensional fermi surface of overdoped la-based
  cuprates}.
\newblock \emph{\bibinfo{journal}{Phys. Rev. Lett.}}
  \textbf{\bibinfo{volume}{121}}, \bibinfo{pages}{077004}
  (\bibinfo{year}{2018}).
\newblock
  \urlprefix\url{https://link.aps.org/doi/10.1103/PhysRevLett.121.077004}.

\end{thebibliography}


\begin{thebibliography}{5}
\expandafter\ifx\csname natexlab\endcsname\relax\def\natexlab#1{#1}\fi
\expandafter\ifx\csname bibnamefont\endcsname\relax
  \def\bibnamefont#1{#1}\fi
\expandafter\ifx\csname bibfnamefont\endcsname\relax
  \def\bibfnamefont#1{#1}\fi
\expandafter\ifx\csname citenamefont\endcsname\relax
  \def\citenamefont#1{#1}\fi
\expandafter\ifx\csname url\endcsname\relax
  \def\url#1{\texttt{#1}}\fi
\expandafter\ifx\csname urlprefix\endcsname\relax\def\urlprefix{URL }\fi
\providecommand{\bibinfo}[2]{#2}
\providecommand{\eprint}[2][]{\url{#2}}

\bibitem[{\citenamefont{Foussats and Greco}(2004)}]{foussats04}
\bibinfo{author}{\bibfnamefont{A.}~\bibnamefont{Foussats}} \bibnamefont{and}
  \bibinfo{author}{\bibfnamefont{A.}~\bibnamefont{Greco}},
  \bibinfo{journal}{Phys. Rev. B} \textbf{\bibinfo{volume}{70}},
  \bibinfo{pages}{205123} (\bibinfo{year}{2004}).

\bibitem[{\citenamefont{Hubbard}(1963)}]{hubbard63}
\bibinfo{author}{\bibfnamefont{J.}~\bibnamefont{Hubbard}},
  \bibinfo{journal}{Proc. R. Soc. London A} \textbf{\bibinfo{volume}{276}},
  \bibinfo{pages}{238} (\bibinfo{year}{1963}).

\bibitem[{\citenamefont{Bejas et~al.}(2017)\citenamefont{Bejas, Yamase, and
  Greco}}]{bejas17}
\bibinfo{author}{\bibfnamefont{M.}~\bibnamefont{Bejas}},
  \bibinfo{author}{\bibfnamefont{H.}~\bibnamefont{Yamase}}, \bibnamefont{and}
  \bibinfo{author}{\bibfnamefont{A.}~\bibnamefont{Greco}},
  \bibinfo{journal}{Phys. Rev. B} \textbf{\bibinfo{volume}{96}},
  \bibinfo{pages}{214513} (\bibinfo{year}{2017}).

\bibitem[{\citenamefont{Bejas et~al.}(2012)\citenamefont{Bejas, Greco, and
  Yamase}}]{bejas12}
\bibinfo{author}{\bibfnamefont{M.}~\bibnamefont{Bejas}},
  \bibinfo{author}{\bibfnamefont{A.}~\bibnamefont{Greco}}, \bibnamefont{and}
  \bibinfo{author}{\bibfnamefont{H.}~\bibnamefont{Yamase}},
  \bibinfo{journal}{Phys. Rev. B} \textbf{\bibinfo{volume}{86}},
  \bibinfo{pages}{224509} (\bibinfo{year}{2012}).

\bibitem[{\citenamefont{Greco et~al.}(2017)\citenamefont{Greco, Yamase, and
  Bejas}}]{greco17}
\bibinfo{author}{\bibfnamefont{A.}~\bibnamefont{Greco}},
  \bibinfo{author}{\bibfnamefont{H.}~\bibnamefont{Yamase}}, \bibnamefont{and}
  \bibinfo{author}{\bibfnamefont{M.}~\bibnamefont{Bejas}}, \bibinfo{journal}{J.
  Phys. Soc. Jpn.} \textbf{\bibinfo{volume}{86}}, \bibinfo{pages}{034706}
  (\bibinfo{year}{2017}).

\end{thebibliography}

\putbib[main_natnew_supp]
\end{bibunit}

\end{document}